\documentclass[10pt, a4]{article}
\pdfoutput=1
\usepackage[a4paper, total={6in, 9in}]{geometry}
\usepackage{fancyhdr}
\usepackage{amssymb}
\usepackage[authoryear]{natbib}
\usepackage{graphicx}
\usepackage{booktabs}
\usepackage{amsmath}
\usepackage{authblk}
\newcommand{\fig}[1]{Fig. #1}
\newcommand{\eqn}[1]{Eq. #1}
\renewcommand{\sec}[1]{Sec. #1}
\renewcommand{\d}{\ensuremath{\mathrm{d}}}
\newcommand{\vol}{\ensuremath{v}}
\newcommand{\cvol}{\ensuremath{V}}
\newcommand{\pd}[2]{\ensuremath{\frac{\partial #1}{\partial #2}}}

\renewcommand{\Re}{\ensuremath{\mathrm{Re}}}
\newcommand{\Ca}{\ensuremath{\mathrm{Ca}}}
\newcommand{\St}{\ensuremath{\mathrm{St}}}
\newcommand{\We}{\ensuremath{\mathrm{We}}}
\newcommand{\e}[1]{\ensuremath{\text{e#1 }}}

\begin{document}
\rhead{Third International Academic Conference of Postgraduates, NUAA}
\lhead{}
\chead{}

\title{Generalized breakup and coalescence models for population balance modelling of liquid-liquid flows}
\author{Marcin Traczyk\thanks{m.traczyk@cranfield.ac.uk}, Robert
Sawko\thanks{r.sawko@cranfield.ac.uk}, Chris Thompson}
%\address{Oil and Gas Engineering Centre, Cranfield University,\\ Cranfield, MK43 0AL, United Kingdom}
\affil{Oil and Gas Engineering Centre, Cranfield University,\\ Cranfield, MK43 0AL, United Kingdom}
%\email{m.traczyk@cranfield.ac.uk}
%\ead{m.traczyk@cranfield.ac.uk}
%\begin{frontmatter}
%\email{m.traczyk@cranfield.ac.uk}

\maketitle

\begin{abstract}

  Population balance framework is a useful tool that can be used to describe
  size distribution of droplets in a liquid-liquid dispersion. Breakup and
  coalescence models provide closures for mathematical formulation of the
  population balance equation (PBE) and are crucial for accurate predictions of
  the mean droplet size in the flow. Number of closures for both breakup and
  coalescence can be identified in the literature and most of them need an
  estimation of model parameters that can differ even by several orders of
  magnitude on a case to case basis. In this paper we review the fundamental
  assumptions and derivation of breakup and coalescence kernels. Subsequently,
  we rigorously apply two-stage optimization over several independent sets of
  experiments in order to identify model parameters. Two-stage identification
  allows us to establish new parametric dependencies valid for experiments that
  vary over large ranges of important non-dimensional groups. This be
  adopted for optimization of parameters in breakup and coalescence models over
  multiple cases and we propose a correlation based on non-dimensional numbers
  that is applicable to number of different flows over wide range of Reynolds
  numbers.

\end{abstract}

\section{Population balance equation (PBE)}
%\nomenclature{$n$}{number density function}

Population balance modelling introduces probabilistic description and allows to
track its evolution over time. First the phase space for a single particle is
defined. The phase space or internal coordinate may contain both dynamic
properties of droplets such as velocity as well as statical properties that
distinguish between diferent species e.g.\ volume, mass etc. Considering general
phase space leads to a multidimensional problem which is not easy to solve
analytically or numerically. Subsequently, a number density function (NDF) is
introduced in order to characterise the population of particles. Finally, a
deterministic equation governing the change in NDF is formulated. A recent
review of the foundations and formulation of population balance equation (PBE)
is given by \citet{solsvik2015}. In the following subsection we describe the
adopted formulation for breakage and coalescence.

\subsection{PBE formulation}
We define the phase space of the PBE problem by particle volume $\vol$ and its
NDF as $n(\vol, t)$. NDF is defined so that $n(\vol, t) \d \vol$ is the number
of drops of size $\left[v, v + \d v \right)$. We assume that the shape of
particles is always spherical. The equation for the evolution of the number
density function is then given by:

\begin{gather}
  \pd{n(\vol, t)}{t}
  + \nabla n(\vol, t)
  =
  B_{br} - D_{br} + B_{coal} - D_{coal},
  \label{eqn:pbe}
\end{gather}
where $D$ and $B$ with appropriate subscripts represent the birth and death
source terms of corresponding processes.
%[ADDED THE CONVECTIVE TERM SINCE WE MENTION LATER OR THAT IT IS APPROXIMATED
%WITH RELAXATION TIME, MT]

At this stage it is worth noting that the formalism we chose is not unique to
breakage and coalescence studies. Some researchers define NDF $\hat{n}(v, t)$
such that $\hat{n}(v, t) \d v$ gives number of drops of size $\left[v, v + \d v
\right)$ per control volume, that is physical space occupied by drops. This
arbitrariness may cause ambiguity in the definition of the source terms. It can
be alleviated by separation of the source terms into breakup and coalescence
rates (which are physical quantities, with units of $s^{-1}$, independent of
mathematical formulation) and a function of the selected NDF formulation.
Therefore, one should derive appropriate rates independently of a chosen
mathematical framework. To illustrate this issue and highlight its importance we
point the reader to a well known correlation for coalescence rate derived by
\citet{coulaloglou1977} which is applicable \textit{only} to population balance
equation for $n(\vol, t)$. If one choses to describe the size of droplets with
their diameter instead of volume, or use $\hat{n}(\vol, t)$ the correlation is
no longer valid.

In this work we choose to follow a formulation where physical coalescence and
breakup rates are independent of NDF formulation or the choice of internal
coordinates. Only the source terms in \eqn{\ref{eqn:pbe}} are affected and
their form is as follows:
\begin{gather}
  D_{br} = g(\vol) n(\vol), \\
  B_{br}
  =
  \int_{\vol}^{\infty} \! \beta(\vol', \vol) g(\vol') n(v') \, d\vol',
  \\
  D_{coal}
  =
  n(\vol) \int_{0}^{\infty} \! Q(\vol, \vol') n(\vol') \, d\vol', \\
  B_{coal}
  =
  \int_{0}^{\vol/2} \! Q(\vol - \vol', \vol') n(\vol') n(\vol - \vol') \,
  d\vol',
\end{gather}
where $g(\vol)$ is breakup rate of droplets of size $\vol$ and $\beta(\vol',
\vol)$ is a probable number of droplets of size $\vol$ created in a breakup of
droplet with volume $\vol'$, often referred to as breakup daughter distribution.
Finally, $Q(\vol, \vol')$ is the coalescence rate between drops of sizes $\vol$
and $\vol'$.

\subsection{Discretisation}

Following \citet{kumar1996} we select $M$ discrete points from the internal
space $\vol_1$, $\vol_2$, \ldots, $\vol_M$ and define the total number of drops
for each segment:
\begin{gather}
  N_i(t)
  =
  \int_{\vol_i}^{\vol_{i+1}} \! n(\vol, t)\, \textrm{d}\vol.
\end{gather}
%\nomenclature{$N_i$}{number of droplets of size $\vol \in \left[\vol_i, \vol_{i+1}\right)$}

In order to obtain $M$ equations for $N_i$ we simply integrate the
equation~(\ref{eqn:pbe}) over intervals $(\vol_i, \vol_{i+1})$. At this point we
face a closure problem as integrals on the right hand side depend on the unknown
function $n$. To resolve it mean value theorem is applies as described in
\citet{kumar1996}. Also, following \citet{hidy1970} we choose uniform
distribution $\vol_i=i \vol_1$ in order to arrive at a simpler form of discrete
equations:
\begin{align}
  \pd{N_i}{t}
  =&
  -N_i(t) g(v_i)
  +
  (v_{i+1} - v_{i})
  \sum_{j=i+1}^{M} \beta(v_i,v_j) g(v_j) N_j
  \nonumber
  \\
  +&
  \sum_{j=1}^{i-1} N_j N_{i-j} Q(v_i, v_{i-j})
  -
  N_i  \sum_{j=1}^M N_j Q(v_i, v_j).
\end{align}

It is the above set of equations which is being solved numerically. The discrete
system requires also boundary conditions which remove breakup death source term
for the smallest class and coalescence source term for the largest class.
Optional source term is added for cases with residence time. The remaining
details of the implementation are postponed to \sec{\ref{sec:methodology}}.

\section{Breakup and coalescence rate}
%\nomenclature{$\epsilon$}{turbulent kinetic energy dissipation rate [$m^2
%\cdot s^{-3}$]}
%\nomenclature{$\varrho_d$}{density of the dispersed phase [$kg \cdot m^{-3}$]}
%\nomenclature{$\varrho_c$}{density of the continuous phase [$kg \cdot m^{-3}$]}
%\nomenclature{$\mu_c$}{dynamic viscosity continuous phase [$Pa \cdot s$]}
%\nomenclature{$\sigma$}{interfacial tension [$N \cdot m^{-1}$]}
%\nomenclature{$C_1$, $C_2$}{non-dimmensional constants of the breakup model}
%\nomenclature{$C_3$, $C_4$}{non-dimmensional constants of the breakup model}
%\nomenclature{$f_{coll}$}{collision frequency [$s^{-1}]$}

A recent review of different breakup models was carried by~\cite{liao2009} and
coalescence models in \citet{liao2010} which provides an overview of existing
models and comparison between them. Vast majority of them include at least one
free parameter that needs to be estimated.

\subsection{Breakup rate}

In this work we utilize the expression derived by \citet{coulaloglou1977} that
assumes the breakup is caused by collisions of drops with tubulent eddies.
They postulated that only energies assosiated with eddies smaller than the
droplet diameter will cause breakup. Other eddies will just carry the droplet
without breaking it. They arrive with the following expression for breakup rate:
\begin{equation}
  g(\vol) = C_1 \vol^{-2/9} \epsilon^{1/3} \exp
  \left(
    - \frac{C_2 \sigma}{\varrho_d \epsilon^{2 / 3} \vol^{5/9}}
  \right)
\end{equation}
where $C_1$ and $C_2$ are empirical constants related to breakup time and the
ratio of surface energy to the mean turbulent kinetic energy of impinging
eddies as noted by \citet{Wang2014}.

\subsection{Coalescence rate}

Coalescence rate can be expressed as a product of collision frequency and
coalescence efficiency, since not all collisions between droplets lead to a
coalescence event (see \citet{liao2010} for more details). Here, we adopt the
expression of \citet{coulaloglou1977} for the coalescence efficiency but we
propose a modified expression for the collision frequency.

We consider here only binary collisions between droplets.  In a frame of
reference associated with one of drops, volume ``swept'' by the other drop in a
time interval $\d t$ is given by the product of interaction surface area equal
to $S_i=\pi (d_1 + d_2) ^ 2 / 4$ and distance traveled by the drop $l = u_{rel}
\d t$.  Where $d_1$ and $d_2$ are diameters of both drops. The ``interaction
volume'' is then given by:

\begin{equation}
  \vol_{i} = l S_i \sim (\vol_1^{1/3} + \vol_2^{1/3})^2 u_{rel} dt
\end{equation}
Probability of a collisions is equal to the ratio of this ``interaction
volume'' to the control volume under consideration $\cvol$. The relative
velocity between drops is taken to be proportional to velocity of the eddy
associated with length scale equal to the droplet diameter. It is therefore
proportional to:
\begin{equation}
  u_{rel} \sim \epsilon^{1/3} \left( \vol_1^{2/9} + \vol_2^{2/9} \right)^{1/2}
\end{equation}

Finally, frequency of collision is proportional to the collision probability
divided by the time interval $\d t$:
\begin{equation}
  f_{coll} = C_3 \frac{\epsilon^{1/3}}{\cvol} (\vol_1^{1/3} + \vol_2^{1/3})^2 
  \left(
    \vol_1^{2/9} + \vol_2^{2/9} 
  \right)^{1/2}
\end{equation}
and the coalescence rate is:
\begin{gather}
  Q(\vol_1, \vol_2) = C_3 \frac{\epsilon^{1/3}}{\cvol} (\vol_1^{1/3} + \vol_2^{1/3})^2 
  \left(
    \vol_1^{2/9} + \vol_2^{2/9} 
  \right)^{1/2}
  \cdot \\ \cdot
  \exp
  \left(
    - \frac{C_4 \mu_c \varrho_c \epsilon}{\sigma^2}
    \left[
      \frac{\vol_1^{1/3}\vol_2^{1/3}}
      {\vol_1^{1/3} + \vol_2^{1/3}}
    \right]
  \right).
\end{gather}
At this stage it is worth noting that in the original publication, the
dependency on $\cvol$ was not included causing it be absorbed into $C_3$
constant during the identification process.

\section{Methodology}
\label{sec:methodology}
The chosen formulation of PBE together with breakup and coalescence models
provide a tool to calculate size distribution of droplets in a liquid-liquid
dispersion, after the model constants are estimated.

Studies where population balance models with simillar closures were performed
many times, including work of \citet{Baldyga1992, Wang2014, maass2012}. In the
great majority the breakup and coalescence model parameters where fitted to
match experimental results from series of simillar experimental cases. When
comparing results from different papers one can see that the parameters
estimated by different authors may vary by several orders of magnitude as
ilustrated by \citet{Wang2014}.

Since our breakup and coalescence closures resemble very much existing formulas
we anticipate large variation of values of $C_1$--$C_4$ parameters on case to
case basis. If any breakup or coalescence model is to be general this large
variations should be addressed and incorporated into mathematical formulation of
the model. Since all model parameters have physical significance and can be
related to physical processes and critical values (like frequency of collisions
or energy distribution of eddies causing breakup) we postulate that they can be
expressed as a function of non-dimensional numbers characterising the flow. Four
most important non-dimensional numbers with respect to population balance
problems are Reynolds number ($\Re$) that describes turbulence levels of the
flow, Weber number ($\We$) that characterises the breakup rate, capillary ($\Ca$)
number governing the film behaviour between colliding drops and Stokes number
($\St$) characterising particle inertia.

Population balance equation is solved for the whole domain, using mean values of
the flow fields, and mean diameter obtained numerically is compared with
experimentally measured values.  We seek parameters $C_1$--$C_4$ that provide
the best fit with experimental data optimized on case-to-case basis, instead of
global optimization to the whole dataset, to identify the relationship between
parameters and non-dimensional numbers characterising each flow. 

\subsection{Solution method}

We solve the discrete form of the population balance equation where the
convective term is replaced with the relaxation term with characteristic
time equal to the mean residence time ($\Theta$) of drops in the domain:
\begin{equation}
  \nabla n(\vol, t) \approx - \frac{1}{\Theta} \left[ n(\vol, t) - n_0(\vol) \right]
\end{equation}
where $n_0$ is the initial size distribution of droplets.

We are interested in a steady state solution to PBE, therefore we calculate the
initial value of source term and advance the solution in time until the value
of source term decreases by three orders of magnitude. This should be a
reasonable indication that the numerical result is close to the steady state
solution.

The code is written in Python programming language, using Scipy package to
provide all necessary components of the solution. The discretized equation is
advanced in time with \texttt{odeint} integration scheme from Scipy. The chosen
integration method solves the first order ordinary differential equation and is
able to switch between stiff (backward differentiation) and non-stiff methods
(Adams method). The optimization is performed on each case to find the best
$C_1$-$C_4$ parameters that fit the experimental results best. Optimization is
performed with the Nelder-Mead downhill simplex algorithm through \texttt{fmin}
function of Scipy package to minimize the function:
\begin{equation}
  \text{error} = \frac{\left( \hat{d} - d_{32}\right)^2}{d_{32}^2}.
  \label{eq:error}
\end{equation}
In the above, $d_{32}$ is Sauter mean diameter measured in experiments and $\hat
d$ is the mean diameter obtained from the distribution function.

The approach poses several problems which we now outline. Firstly, during the
optimization process model parameters being optimized often take
values that cause the distribution function to fall outside of the discretized
domain in the phase space. This result in numerical issues with integration and
non-physical result. To avoid unnecessary computational effort we check the
total mass every five iterations and abort the calculation if it falls outside
imposed bounds. Total mass can be estimated from the first moment of the
distribution function. 

\begin{figure}
  \includegraphics[width=0.9\textwidth]{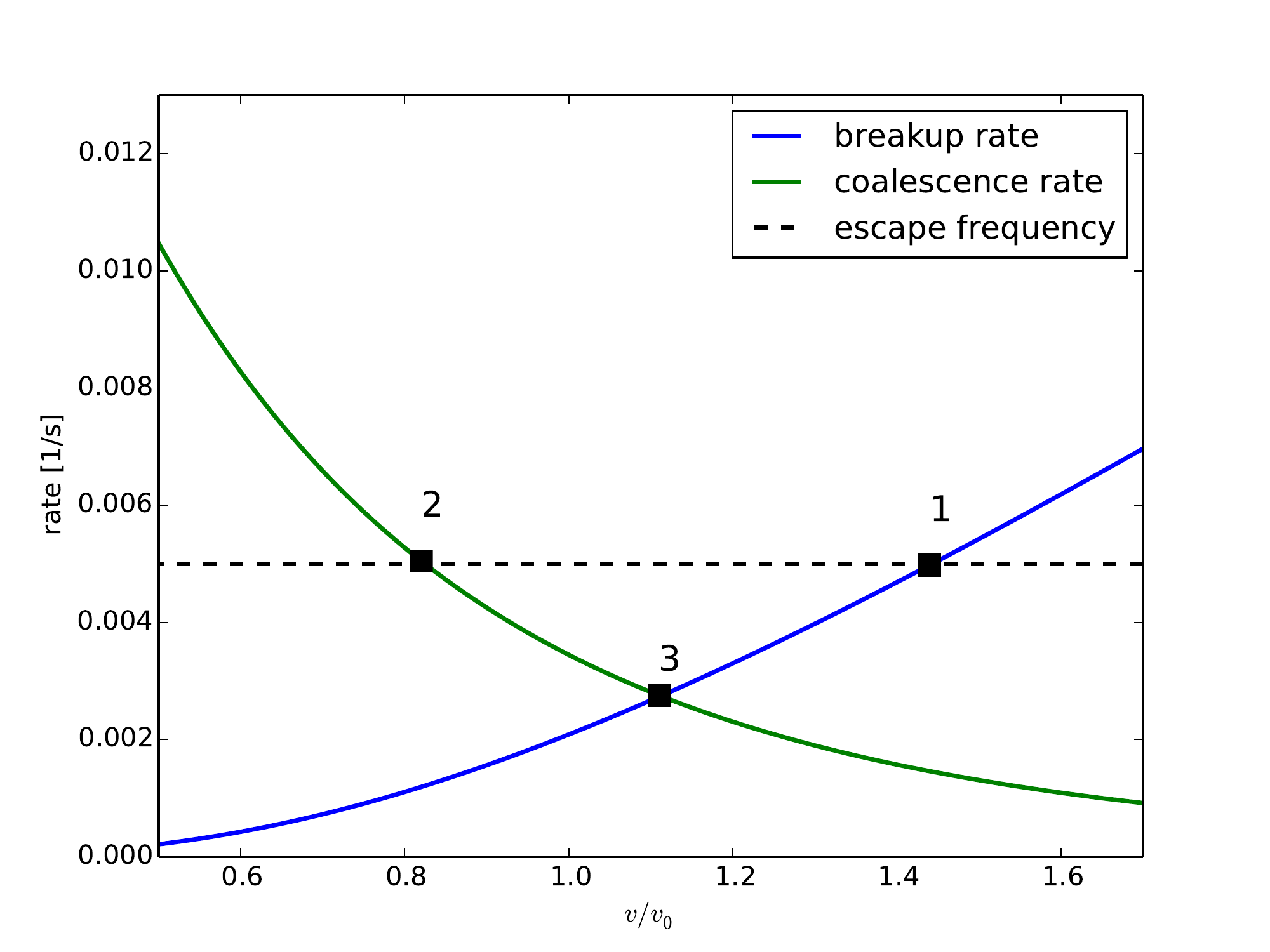}
  \caption{Stable solution to optimization of PBE source terms are located
    around points indicated with squares. \label{fig:minima}. `1' corresponds to
  pure breakup solution, `2' is pure coalescence, `3' is balanced solution.}
\end{figure}

There are three possible stable solutions to a population balance problem as
illustrated on \fig{\ref{fig:minima}}:
\begin{enumerate}
  \item pure breakup - balance between breakup rate and escape frequency; no
    coalescence
  \item pure coalescence - balance between coalescence rate and escape frequency; no
    breakup
  \item balanced solution - balance between coalescence rate, breakup rate and
    escape frequency.
\end{enumerate}
In cases with finite residence time, the optimization algorithm, if given the
initial guess that results in smaller mean diameter than the experimental one,
tends to find a solution to the problem that is a ``pure coalescence'' solution
with almost no breakup. When the initial guess has the mean diameter larger than
the expected value, it finds a ``breakup only'' solution to the problem. To
avoid pushing the system towards unbalanced solutions we calculate the error as
a sum of two cases: one with initial guess with smaller mean diameter and one
with bigger mean diameter. In this way the only mathematical solution to the
problem are balanced coalescence and breakup terms. 

\begin{figure}
  \begin{center}
    \includegraphics[width=0.49\textwidth]{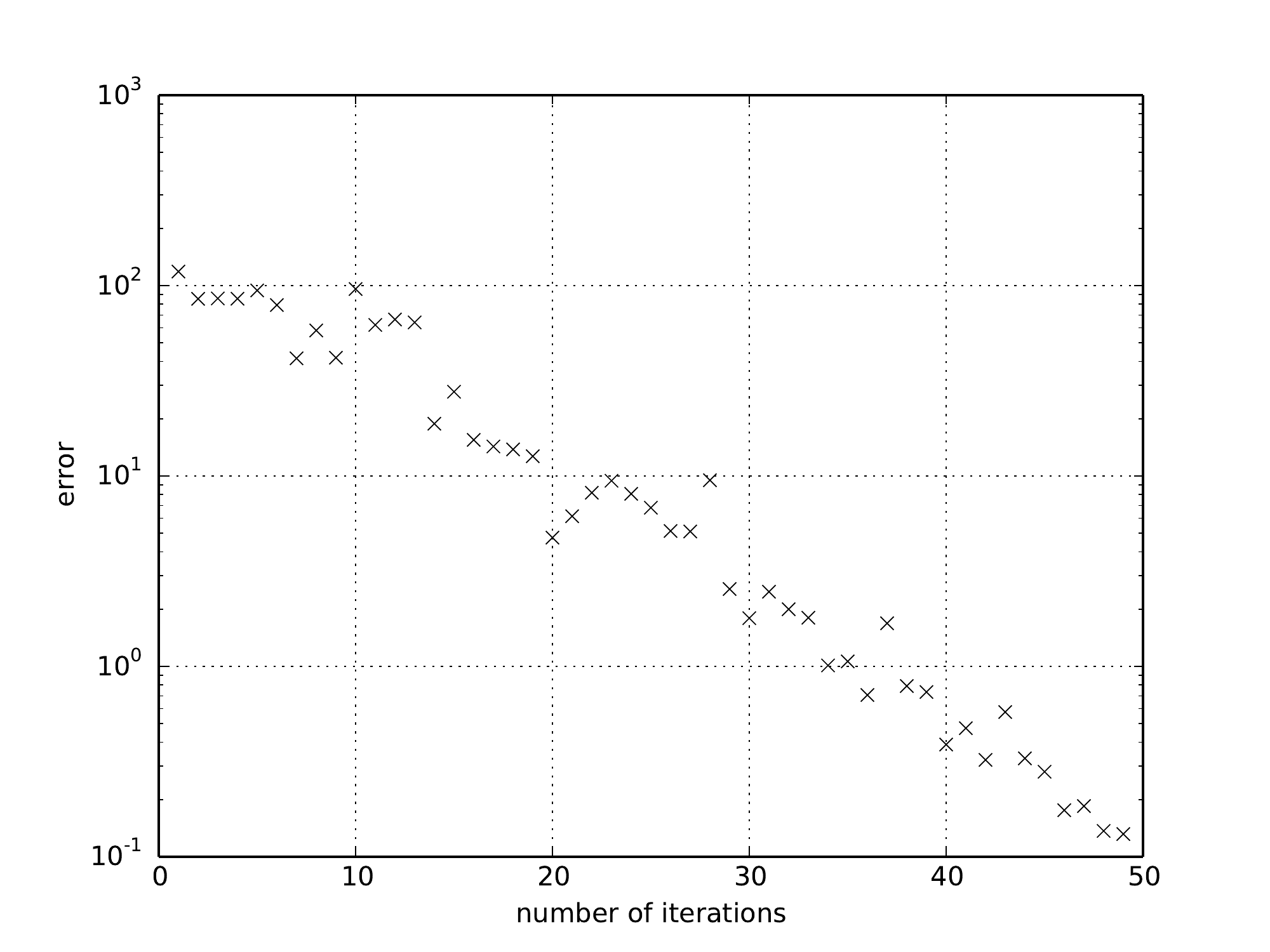}
    \includegraphics[width=0.49\textwidth]{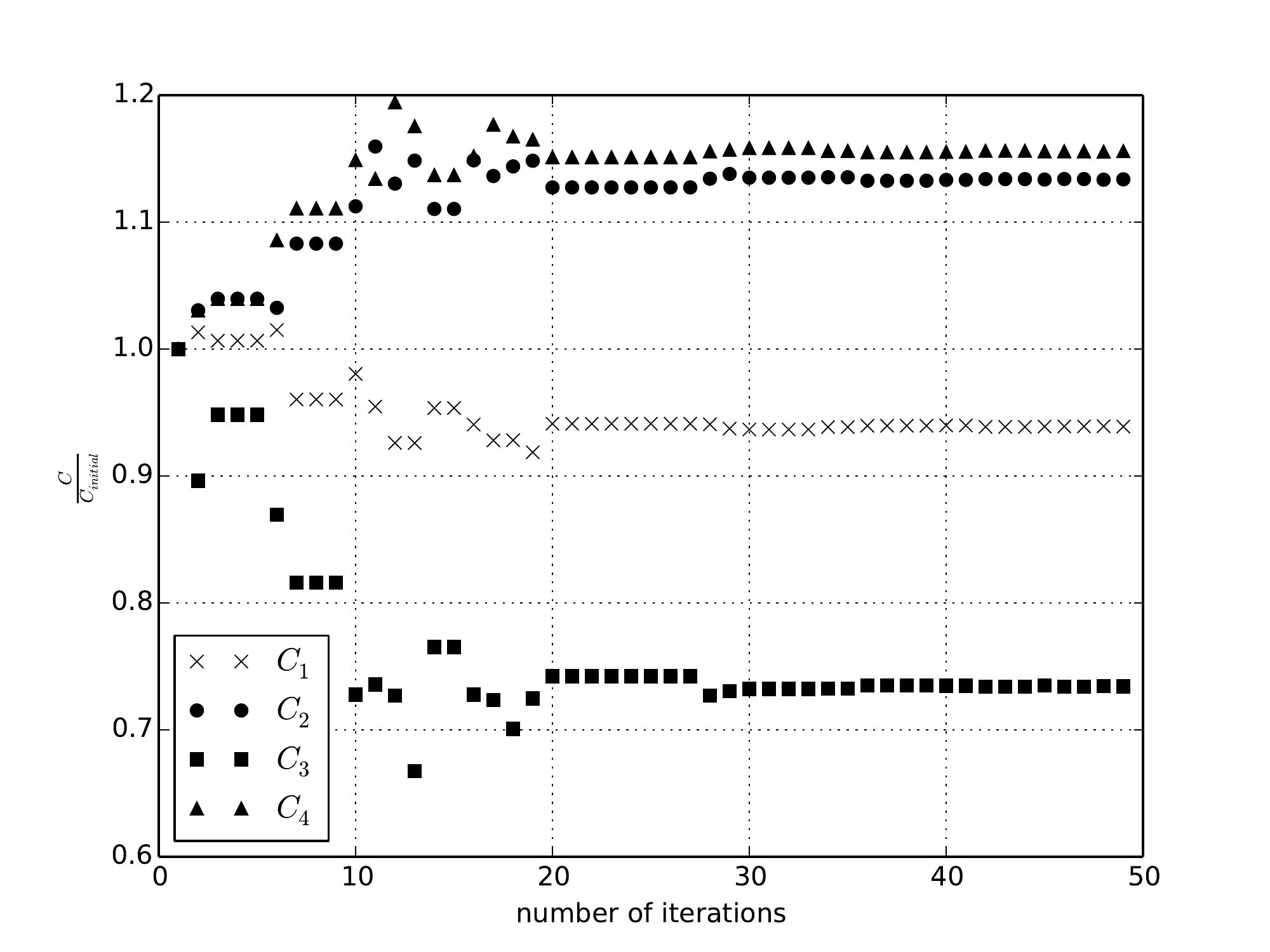}
    \caption{Rate of convergence of optimization process. Left-hand side plot
      shows error as defined by \eqn{\ref{eq:error} as a function of number of
      iterations. The second plot shows relative change in $C_1$--$C_4$
      parameters being optimized.}
      \label{fig:convergence}
    }
  \end{center}
\end{figure}

All the above result in an efficient algorithm to optimize steady state solution
to population balance equation. Typical convergence rate of optimization is illustrated
on \fig{\ref{fig:convergence}}.

\subsection{Selected cases}

%\paragraph{Orifice flow from \citet{galinat2005}} Breakup dominated case with
%no reported coalescence. The experimental setup consists of a $3cm$ diameter
%pipe with a $5mm$ thick, concentric orifice that restricts half of the pipe
%diameter. The liquid-liquid system used is tap water for the continuous phase
%and $n$-heptane for the dispersed phase. For the detailed description of the
%experiment we refer the reader to the original paper of \citet{galinat2005}.

%The numerical studies in this work are conducted using the mean quantities
%characterising the flow in the orifice section since the breakup occurs only in
%the orifice vicinity. For estimate of turbulent kinetic energy dissipation in
%the orifice we use the expression provided by \citet{galinat2005} based on the
%pressure drop across the orifice:
%\begin{equation}
  %\epsilon = \frac{1}{\rho_c} \frac{\Delta P_{max} U}{2D}
  %\left(\frac{1}{\beta^{2} - 1}\right)
%\end{equation}
%where $U$ is the mean flow velocity in the pipe, $D$ is the pipe diameter
%and $\beta$ is the restriction ratio equal to $0.5$.
%Because droplet breakup is caused by the presence of orifice, we calculate
%non-dimensional number based on the flow parameters in the orifice:
%\begin{gather}
  %\Re = \frac{\varrho_c U D }{\mu_c \beta} \\
  %\Ca = \frac{\mu_c U}{\sigma \beta^2} \\
  %\St = \frac{2 \varrho_c}{2 \varrho_d + \varrho_c} \frac{2 d_{32}^2}{9 D^2
  %\beta^2} \Re
%\end{gather}

\paragraph{Test case}

\begin{figure}
  \includegraphics[width=0.9\textwidth]{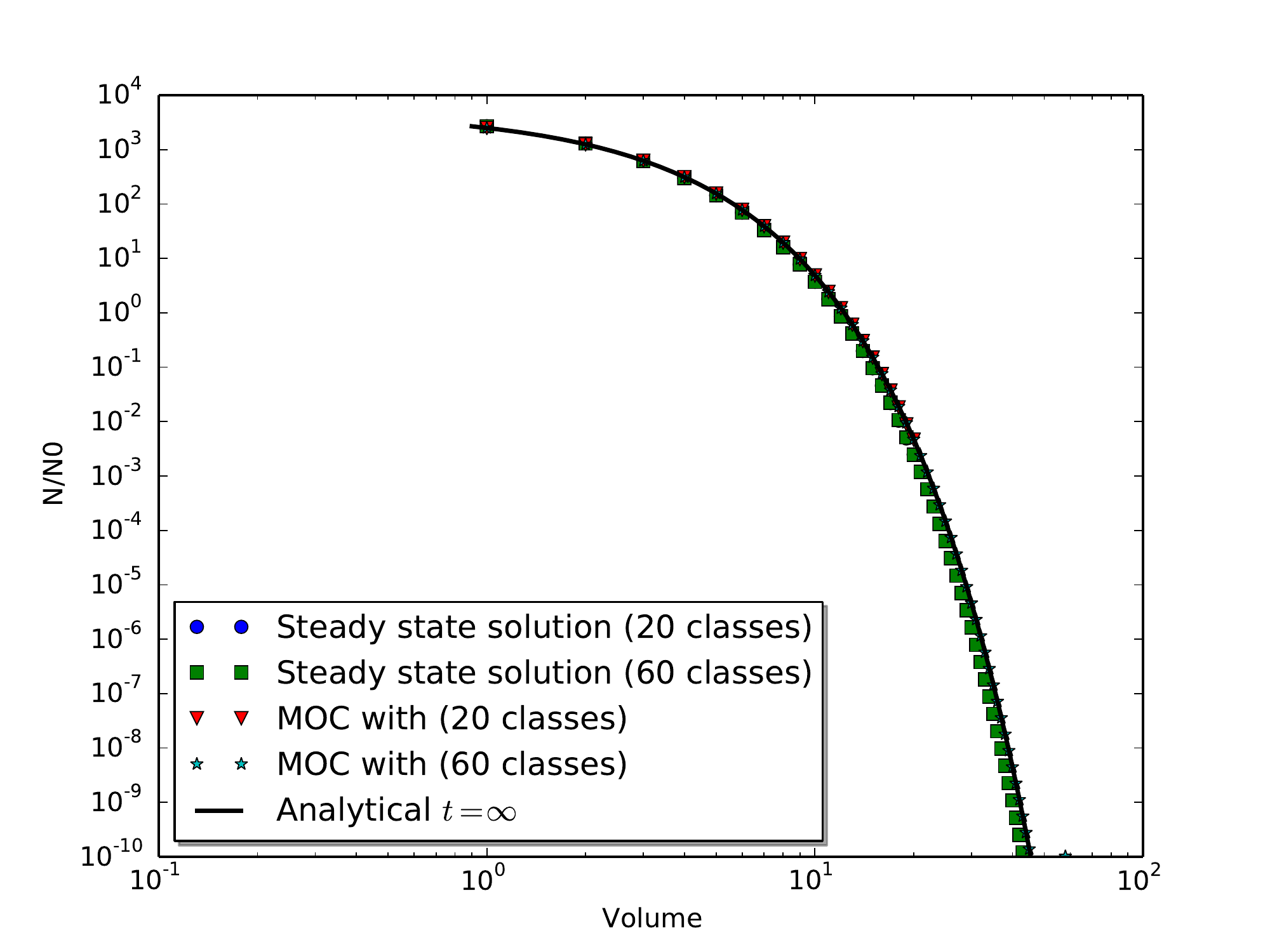}
  \caption{Comparison of steady stat solution, transient solution and
  analytical result. 
  \label{fig:test}.}
\end{figure}

To validate the methodology and implementation a test case with known
analytical solution has been chosen. The selected case is simultaneous breakup
and coalescence of polymers from \citet{blatz1945}. Comparison between solution
obtained with transient PBE solver, steady state solution iterated until source
term value dropped to $1\e{-03}$ of initial value and analytical solution can be
seen on \fig{\ref{fig:test}}. The chosen threshold of $1\e{-03}$ is a reasonably
good indication of the steady state and an idealized test case can be
reproduced with as few as twenty size classes.

\paragraph{Horizontal pipe flow from \citet{simmons2001}} Authors report number
of experiments on both upward and horizontal flows of kerosene (continuous
phase) and potassium carbonate solution in a $63mm$ pipe. In the paper Sauter
mean diameters are reported in three different position in the horizontal
systems: bottom, center-line, high position. We choose only one of their cases,
with sufficiently high Reynolds number to produce dispersion that is uniform
enough to have almost the same mean diameters measured in three different
positions. Therefore we take a case with mean flow velocity of $2.71m/s$ and
$11.7\%$ of the dispersed phase by volume. For more details we once again refer
the reader to the original paper of \citet{simmons2001}, and for even more
detailed description to the thesis of \citet{simmonsPhD}.

\paragraph{Continuous flow stirred tank from \citet{coulaloglou1977}} 
Experiments were performed in a continuous flow $12l$ baffled stirred tank with a
$10cm$ turbine impeller. The liquid-liquid system used was water for the
continuous phase and kerosene-dichlorobenzene as the dispersed phase.
Non-dimensional numbers governing the flow were calculated based on impeller
geometry (as used to describe similar systems for example by \cite{wang1986,
coulaloglou1977}):
\begin{gather}
  \Re = \frac{N_{i} D_{i}^2}{\mu_c \varrho_c} \\
  \Ca = \frac{\mu_d N_i D_i}{\sigma} 
  \left(
    \frac{\varrho_c}{\varrho_d}
  \right)^{1/2} \\
  \St = \frac{2 \varrho_c}{2 \varrho_d + \varrho_c} \frac{2 d_{32}^2}{9 D_i^2
  \beta^2} \Re
\end{gather}
with $N_i$ and $D_i$ being, accordingly, impeller speed (in revolutions per
second) and impeller diameter.
Turbulent kinetic energy dissipation is calculated in the same way as in
\citet{coulaloglou1977}, through formula:
\begin{equation}
  \epsilon = 0.407a N_i^3 D_i^2
\end{equation}

\paragraph{Horizontal pipe flow from \citet{angeli2000}} Paper investigates the
effect of the pipe material on the droplet size distribution. For the
validation we selected six oil(continuous)-water(dispersed) flows in a $24mm$
diameter acrylic pipe.

\paragraph{Horizontal pipe flow from \citet{karabelas1978}} Author reports
oil-water flows in $5.04cm$ diameter pipe at various concentrations from which
we select 7 cases.

\subsection{Initial conditions}
In the fully developed pipe flows the convective term is not taken into account since
boundary condition should not effect the result. In cases taken from
\citet{coulaloglou1977} we take the residence time to be $10min$, exactly the
same as reported by the authors. 

As an initial distribution for every simulation we take a Gaussian-type
function:
\begin{equation}
  n_0 = \frac{\alpha \cvol}{\vol_0} \frac{1}{\sqrt{2\pi} \sigma_0}
  \exp \left(
    \frac{(\vol - \vol_0)^2}{2 \sigma_0^2}
  \right)
\end{equation}
with $\vol_0$ and $\sigma_0$ being initial mean volume of drops and their
standard deviation and $\alpha$ being volumetric fraction of the dispersed phase
in the domain.
%In the orifice flow, we take the residence
%time to be twice the time needed for the flow ($U_{or} =
%\frac{U} / \beta^2 = 4U$) to travel the distance equal to the orifice thickness
%($5mm$). This is somewhat arbitrary but it might be justified by the fact that
%most of the breakup reported by \citet{galinat2005} took place just after the
%orifice.

\begin{table}
  \begin{center}
    \begin{tabular}{c c}
      \toprule
      case & residence time \\
      \midrule
      \citet{simmons2001} & $\infty$ \\
      \citet{angeli2000} & $\infty$ \\
      \citet{karabelas1978} & $\infty$ \\
      \citet{coulaloglou1977} & $10\text{min}$ \\
      %\citet{galinat2005} & $\frac{0.04}{U}$ \\
      \bottomrule
    \end{tabular}
    \caption{Residence times used in simulations. Infinite time corresponds to
    lack of inflow/outflow terms.}
  \end{center}
\label{tab:coeffs}
\end{table}

\section{Results}
\label{sec:results}
\subsection{Parameter dependency on non-dimensional numbers}

\begin{figure}
  \center{
  \includegraphics[width=0.49\textwidth]{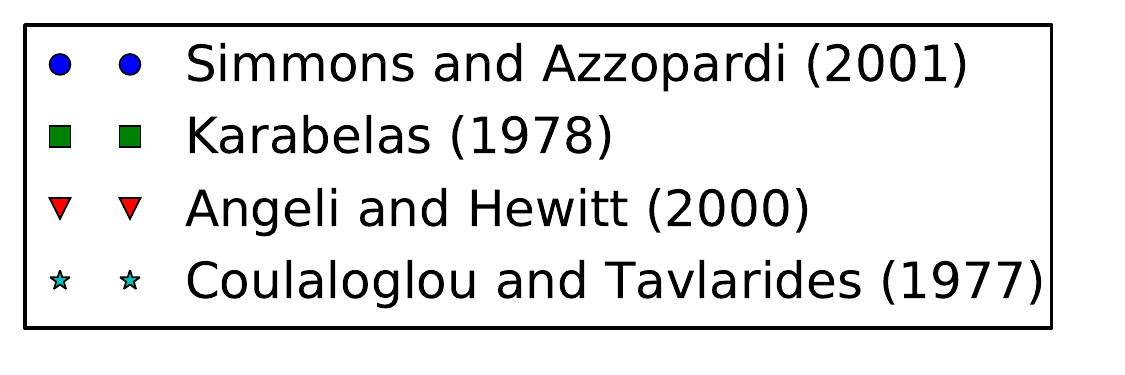}} \\
  \includegraphics[width=0.49\textwidth]{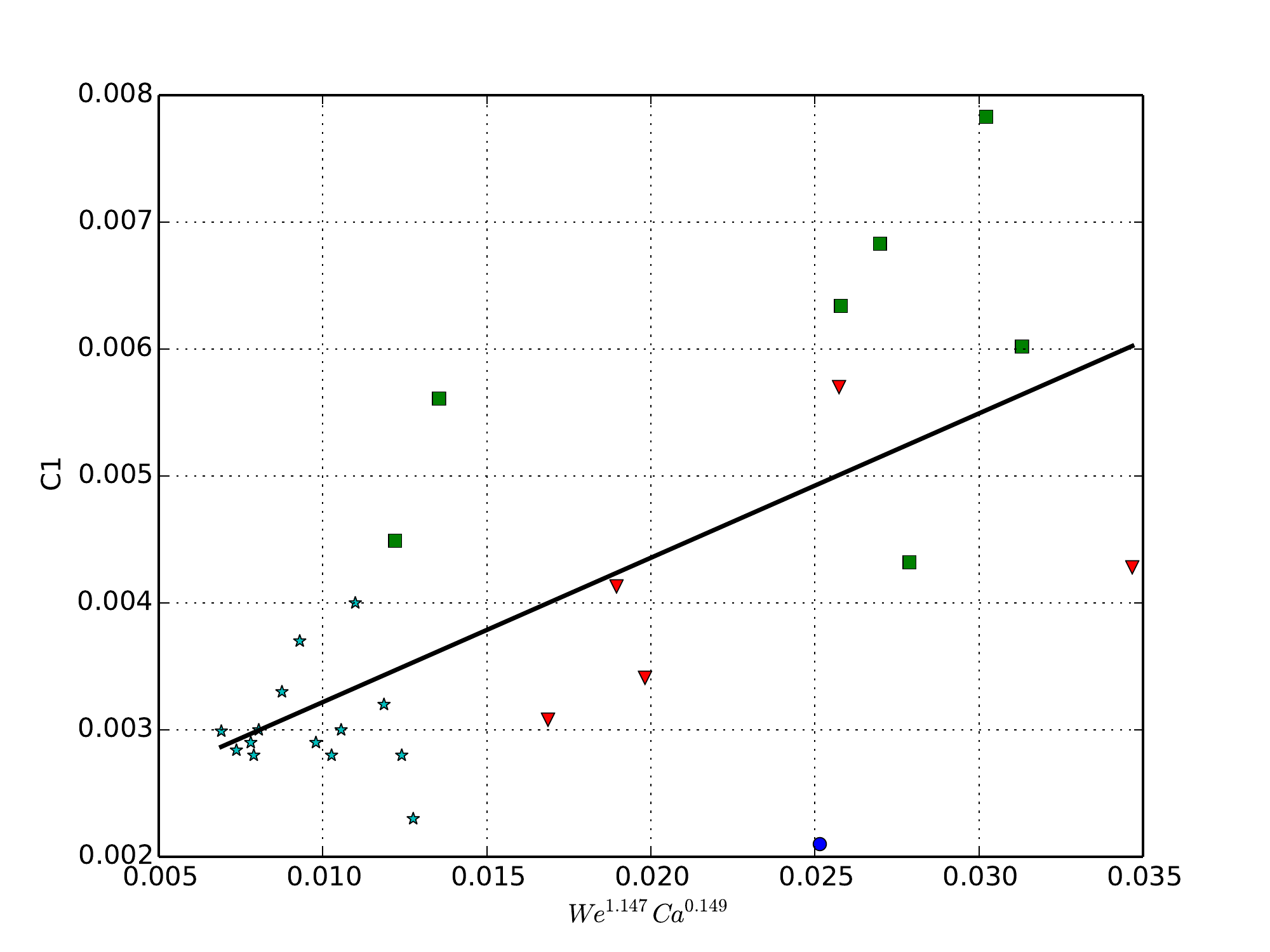}
  \includegraphics[width=0.49\textwidth]{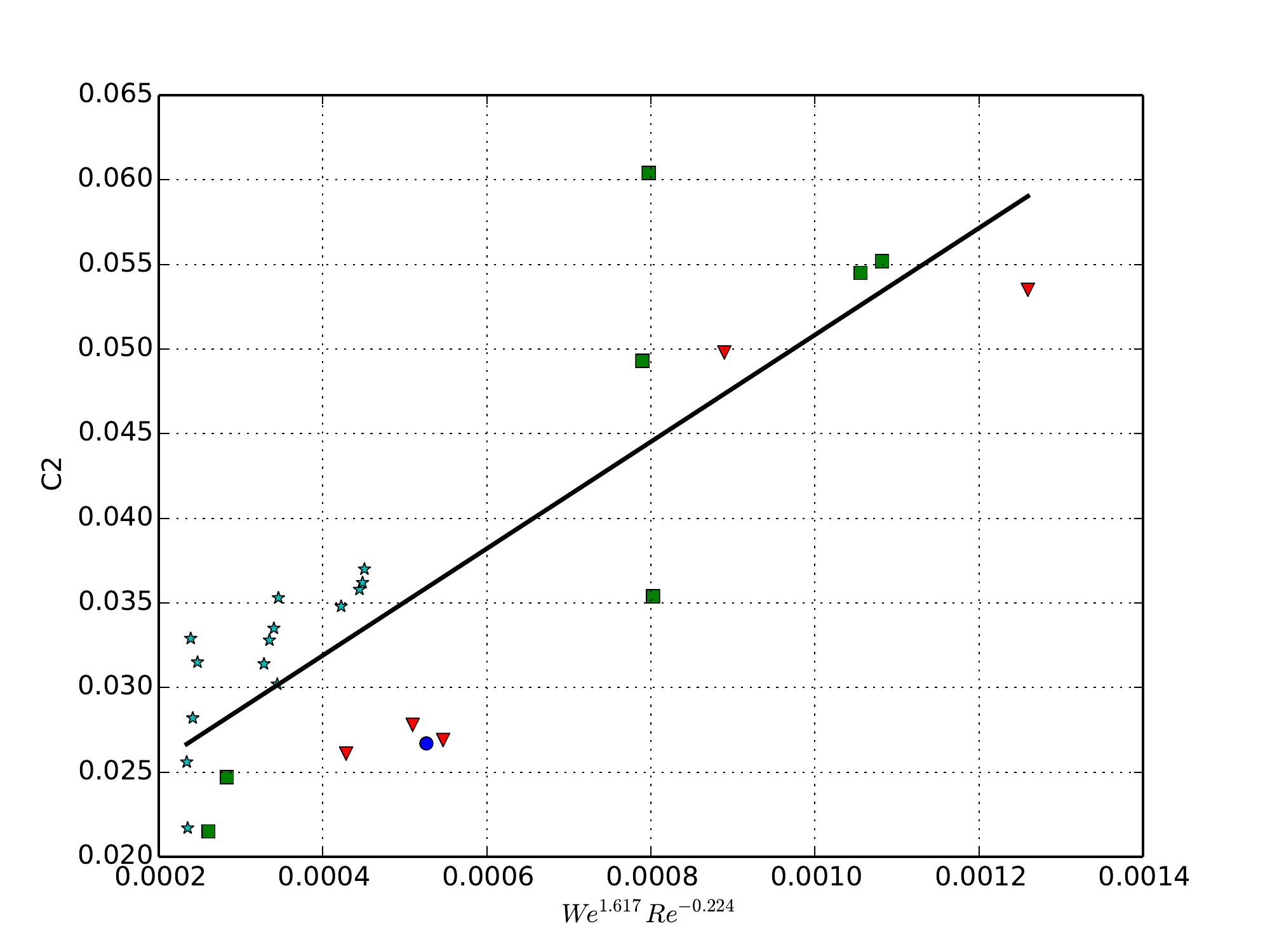}
  \\
  \includegraphics[width=0.49\textwidth]{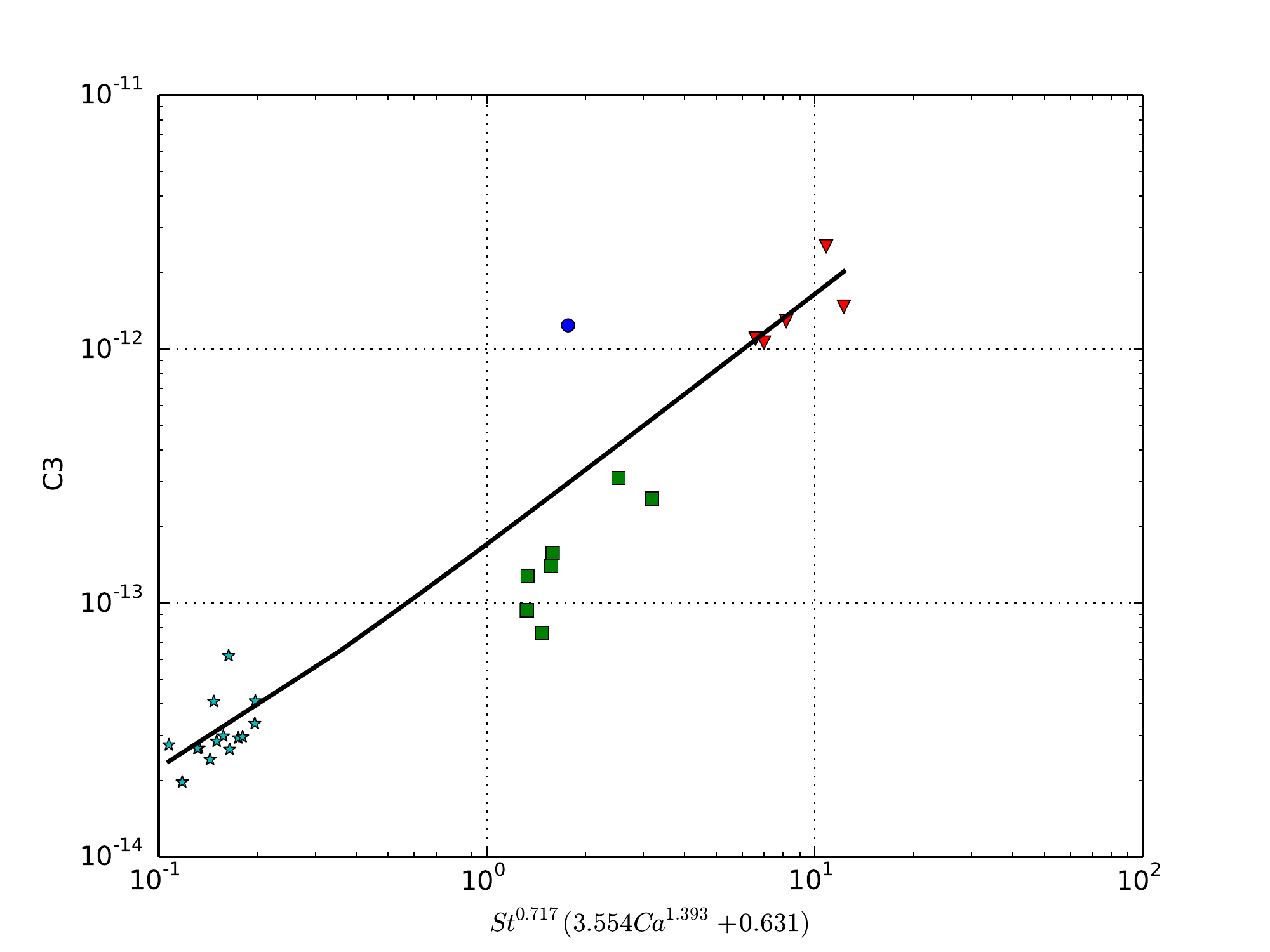}
  \includegraphics[width=0.49\textwidth]{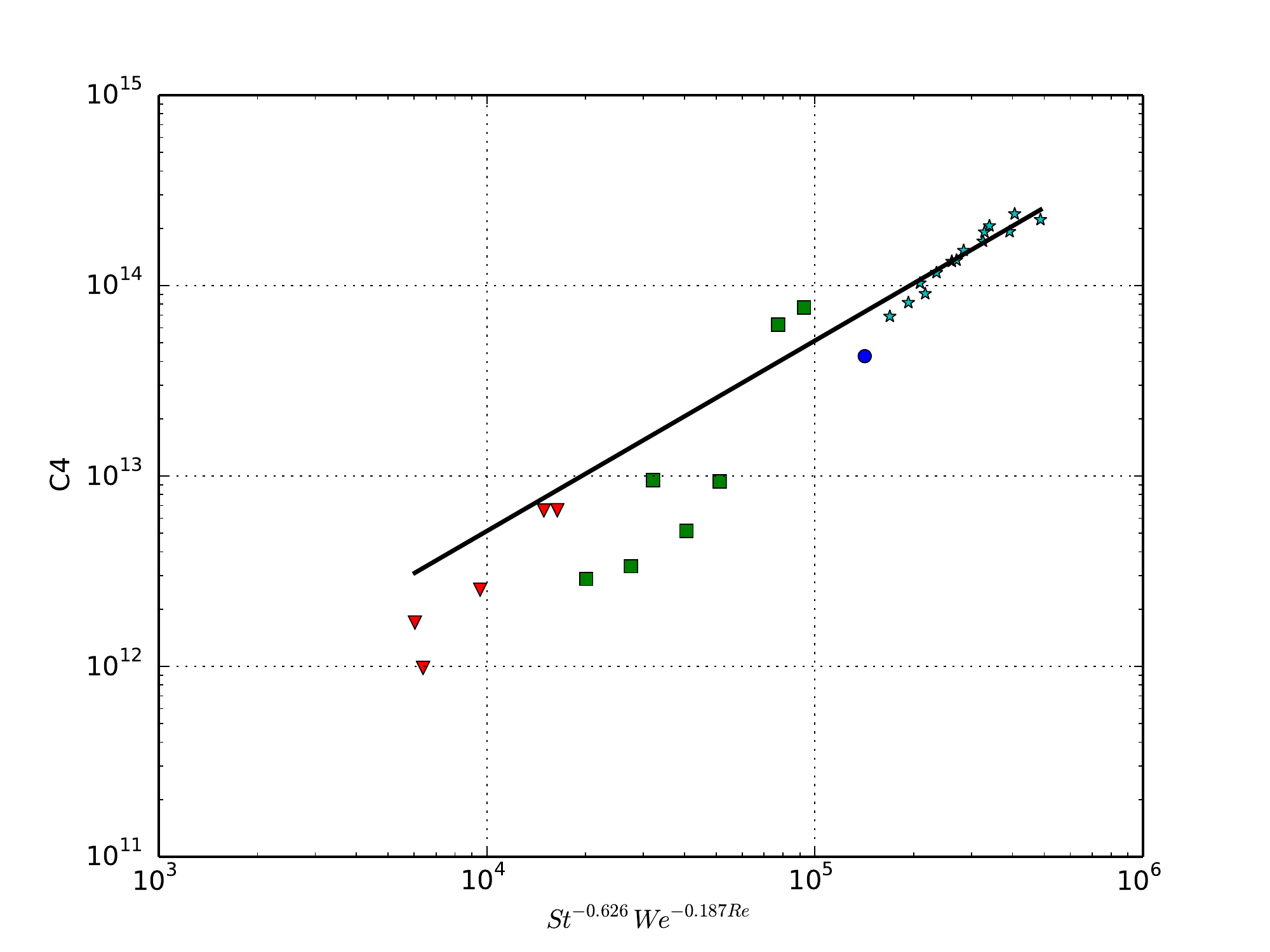}
  \caption{Model parameters together with proposed correlations. \label{fig:C}}
\end{figure}

\begin{figure}
  \includegraphics[width=0.9\textwidth]{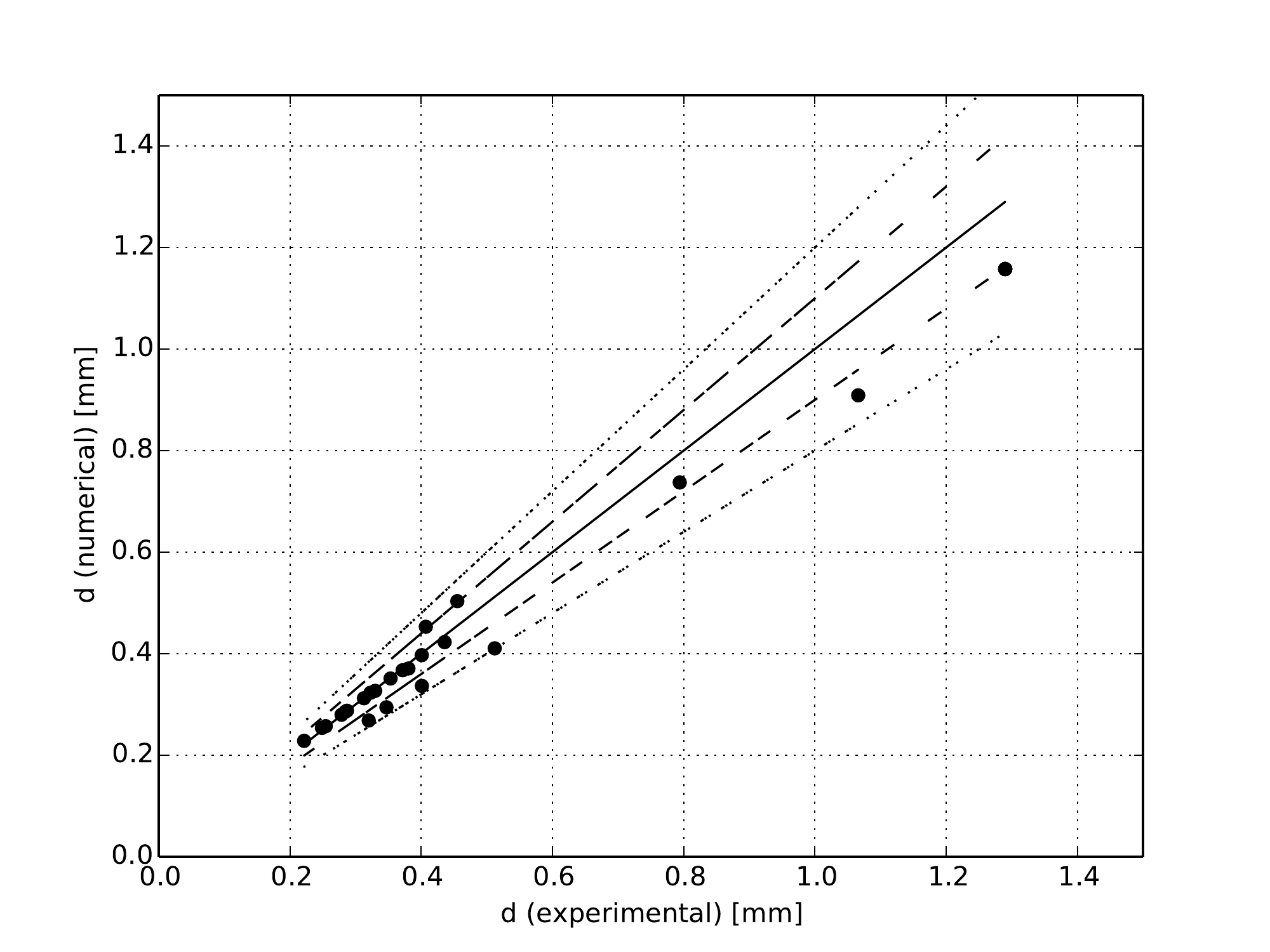}
  \caption{Proposed model predictions versus experimental Sauter mean diameter.
  Continuous line corresponds to an exact match, dashed lines represent 10\% and
  20\% error accordingly. \label{fig:comparison}}
\end{figure}

Attempt to find the functional dependency on $\Re$, $\St$, $\We$ and $\Ca$ was
perform after identification of $C_1$ -- $C_4$ parameters for each case. The
$C_3$ and $C_4$ parameters from coalescence model varied in the selected cases
by two orders of magnitude and it is clear that the parameters can be expressed
as a function of all important non-dimensional numbers. Their values change
monotonically with increasing $\Re$, $\St$, $\We$ and $\Ca$. We propose easy to
use correlations expressing the value of coalescence model parameters as a
product of non-dimensional groups raised to a certain powers. From least square
fitting we find that:

\begin{gather}
  C_3 =  1.6429\e{-13} \St^{0.7174} \left(3.5541\Ca^{1.3928} + 0.6311 \right) \\
  C_4 = 5.137\e{08} \St^{-0.6258} \We^{-0.1874} \Re
\end{gather}

The breakup model parameters have not displayed such a clear dependency on chosen
non-dimensional numbers since their variation was much smaller than coalescence
parameters. There are at least four possible explanations for this
situation:
\begin{enumerate}
  \item functional dependency of $C_1$ and $C_2$ on flow fields is more
    complicated than assumed dependency on non-dimensional groups
  \item the parameters are constant and the results we obtained can be explained
    as noisy data scattered around the true value
  \item current population balance models do not capture all the essential
    physics of the problem: either solving population balance equation solved
    for the whole domain is not accurate enough and using mean dissipation in
    the domain is not representative or used correlations for breakup and
    coalescence rates do not take into account all sources of breakup and
    coalescence events
  \item the dependency of $C_1$ and $C_2$ on the mean flow values is small and
    uncertainties involved in the approach taken by us prohibit us from
    identifying it
\end{enumerate}

We choose to treat $C_1$ and $C_2$ as parameters weakly dependent on several
non-dimensional groups that provided best fit with experimental data:
\begin{gather}
  C_1 = 0.1137 \We^{1.1467} \Ca^{0.1491} + 1.0814\e{-03} \\
  C_2 = 31.56 \We^{1.6170} \Re^{-0.2239} + 0.01927
\end{gather}

Above expression were used to calculate $C_1$ -- $C_4$ values for the final set
of simulations and the comparison of obtained mean diameter and experimental
values from all 27 simulations are plotted on \fig{\ref{fig:comparison}}. We find
that most of the results are within 10\% error from experimental values and all
results fall withing $\pm$20\% error band. The results obtained prove that the
established model can be successfully applied to number of liquid-liquid flows
across a wide range of non-dimensional parameters.

\subsection{Recommendations for the future work}

Since the physical mechanisms of breakup and coalescence in bubbly flows of
other types of dispersed flows are not different from the ones in liquid-liquid
flow, similar optimization procedure can be applied to them. It might be
possible to find an even more generalized expressions that can be applied for
wider range of flows.

Incorporating larger number of experimental cases in the optimization process
should result in more reliable correlations also applying the results of this
work in a full three-dimensional CFD simulation should help in gaining more
confidence in obtained results.

%For the purpose of expanding the work undertaken here we make all the
%optimization scripts available in the public domain at www\dots.

\section{Acknowledgements}

\begin{raggedright}
This work has been undertaken within the Consortium on Transient and Complex
Multiphase Flows and Flow Assurance (TMF). The Authors wish to acknowledge the
contributions made to this project by the UK Engineering and Physical Sciences
Research Council (EPSRC) and the following: - ASCOMP, BPExploration; Cameron
Technology \& Development; CD-adapco; Chevron; KBC (FEESA); FMC Technologies;
INTECSEA; Institutt for Energiteknikk (IFE); Kongsberg Oil \& Gas Technologies;
MSi Kenny; Petrobras; Schlumberger Information Solutions; Shell; SINTEF; Statoil
and TOTAL. The Authors wish to express their sincere gratitude for this support.
\end{raggedright}

%% The Appendices part is started with the command \appendix;
%% appendix sections are then done as normal sections
%% \appendix

%% \section{}
%% \label{}

%% If you have bibdatabase file and want bibtex to generate the
%% bibitems, please use
%%

%\printnomenclature
\bibliographystyle{abbrvnat} 
%\bibstyle
\bibliography{ref}

%% else use the following coding to input the bibitems directly in the
%% TeX file.

%\begin{thebibliography}{00}

%% \bibitem[Author(year)]{label}
%% Text of bibliographic item

%\bibitem[ ()]{}

%\end{thebibliography}
\end{document}